\documentclass[prl,aps,letterpaper,floatfix,superscriptaddress,nofootinbib,preprintnumbers,twocolumn,10pt]{revtex4-1}

\usepackage{dcolumn}
\usepackage{bm}
\usepackage[dvips]{graphicx}
\usepackage{epsfig}
\usepackage{amsmath, amsfonts, amssymb, slashed}

\newcommand{\be}{\begin{equation}}
\newcommand{\ee}{\end{equation}}
\newcommand{\bee}{\begin{equation*}}
\newcommand{\eee}{\end{equation*}}

\usepackage{color}

\newcommand{\bea}{\begin{eqnarray}}  
\newcommand{\eea}{\end{eqnarray}}

\usepackage{feynmp}
\usepackage{tikz}
\begin{document}

\title{Drell-Yan Constraints on New Electroweak States:\\
LHC as a $p p \to \ell^+\ell^-$ Precision Machine
}

\author{Christian Gross}
\author{Oleg Lebedev}
\affiliation{Department of Physics and Helsinki Institute of Physics,
Gustaf H\"allstr\"omin katu 2, FIN-00014 University of Helsinki, Finland}
\author{Jose~Miguel~No}
\affiliation{Department of Physics and Astronomy, University of Sussex, Brighton 
BN1 9QH, United Kingdom}


\begin{abstract}
The Standard Model (SM) extensions with vector--like states which 
have either zero hypercharge or zero weak isospin are rather poorly 
constrained by the electroweak precision measurements.
Such new states would however modify the 
running of the gauge couplings at high energies. 
As a result, 
 the Drell--Yan process $p p \to \ell^+\ell^-$ 
at the LHC places useful constraints on these models. 
The relevant observables include both the di-lepton invariant mass distribution 
$M_{\ell\ell}$ and the forward-backward asymmetry $A_{\mathrm{FB}}$.
 We find that the LHC Run 1   data  and  the initial data from   Run 2  surpass the sensitivity of 
LEP and already put meaningful constraints on the existence of such particles, which will become progressively stronger with more data.
\end{abstract}

\maketitle
\section*{Introduction}

\vspace{-3mm}

The existence of new particles  at the electroweak  (EW) scale  can be motivated by various considerations including the hierarchy 
problem and the WIMP solution to the dark matter puzzle. Many such models   are constrained by measurements of EW precision observables (EWPO)
(see e.g.~\cite{Barbieri:2004qk}). However, if the new states have either zero hypercharge or zero weak isospin, such constraints become 
very weak. Direct searches for these  states at the LHC are highly model dependent, especially if the new fermions carry no color.  
The current bounds can be as low as $M_f \sim 100-200$ GeV depending on the specifics of the model~\cite{Achard:2001qw, Falkowski:2013jya, Aad:2015dha,Djouadi:2016eyy}.

On the other hand, the  presence of  new EW states affects the renormalization group (RG) running of the SM gauge couplings at high energies~\cite{Alves:2014cda},
which could be probed at the LHC via precise measurements of 
the Drell-Yan production rate~\cite{Drell:1970wh}.
In this work, we focus on the process $pp \to \gamma^*/Z^* \to \ell^+ \ell^-$  at large di-lepton invariant masses.
We find that the 
 LHC Run 1 data at $\sqrt{s} = 8$ TeV greatly surpass the sensitivity of EWPO and   provide non-trivial constraints on the existence 
of   such electroweak states. We also show that the LHC Run 2 data will significantly improve the LHC 8 TeV bounds. 
Moreover, unlike the bounds from direct searches, 
these limits are insensitive to the details of the spectrum and couplings among the new states.

The Drell-Yan process $pp \to \gamma^*/Z^* \to \ell^+ \ell^-$ at high di-lepton invariant mass $M_{\ell\ell}$ combines
a significant cross section $\sigma$ with good control of theoretical and experimental uncertainties,  allowing for precision measurements 
of the SM EW gauge couplings~\cite{Alves:2014cda}.  Here we explore two observables: the di-lepton
invariant mass distribution $d\sigma/dM_{\ell\ell}$ and the di-lepton
forward-backward asymmetry $A_{\mathrm{FB}}$. These observables are complementary, being sensitive to different 
combinations of the gauge couplings. We stress that $d\sigma/dM_{\ell\ell}$ data from the 8 TeV LHC Run already
place interesting constraints on the scenarios at hand, while   $A_{\mathrm{FB}}$ will become important with more data.  

\vspace{-3mm}

\section{ Drell-Yan Signatures of New Physics}

\vspace{-2mm}

In order to assess the sensitivity of $d\sigma/dM_{\ell\ell}$ and $A_{\mathrm{FB}}$ to new physics, let us
extend the SM by a set of vector-like fermions with mass $M_f$. To  isolate their effect 
on $\alpha_Y=g_{Y}^2/4\pi$ and $\alpha_2= g_{2}^2/4\pi$, respectively, we assume that
the new fermions are either $N$ copies of $SU(2)_L$ singlets with hypercharge $Y_f$, or $N_d$ copies of $SU(2)_L$ doublets with zero 
hypercharge (here $N$, $N_d$  include the color factor for the case of vector-like quarks). 
Such SM extensions are very weakly constrained by EWPO. The relevant EW parameters are
$\mathrm{Y} =  N \,Y_f^2 \times \alpha_Y \, m_W^2 / (15 \pi M_f^2)$ and $\mathrm{W} = N_d  \times \alpha_2 \, m_W^2 / (30 \pi M_f^2)$,
which are constrained at the level of a few per mille~\cite{Barbieri:2004qk}, rendering the EWPO constraints essentially irrelevant
for $M_f > 400$ GeV.

The new states contribute to the RG running of the EW couplings by changing the SM beta functions by
\begin{equation}
\Delta \beta_Y = \frac{g_Y^3}{16 \pi^2}\; \frac{4 N \,Y_f^2}{3} \;\;\;\;, \;\;\;\;
\Delta \beta_2 = \frac{g_2^3}{16 \pi^2}\; \frac{2 N_d}{3}
\end{equation}
for scales $\mu > M_f$. In what follows, we focus on the leading log corrections to $\alpha_i$ at high 
energies $\mu \gg M_Z$, $\delta \alpha_i(\mu) \equiv \alpha_i(\mu) - \alpha^{\mathrm{SM}}_i (\mu)$. 



\vspace{2mm}

\noindent \textbf{Di-Lepton Invariant Mass Distribution}. 
The Born level $pp \rightarrow \gamma^*/Z^* \rightarrow \ell^+ \ell^-$ 
differential cross section is given by 
\bea
\label{eq1}
\frac{d \sigma} {d M_{\ell\ell}} &=& \frac{\pi}{648\, M_{\ell\ell}^3} \sum_q C_q(\alpha_Y,\alpha_2,M_{\ell\ell})\nonumber\\ 
&\times &\int^{\ln \frac{\sqrt{s}}{M_{\ell\ell}}}_{-\ln \frac{\sqrt{s}}{M_{\ell\ell}}} dy \left[   x_1 f_q (x_1) \,x_2 f_{\bar{q}}(x_2) \right],
\eea
where $M_{\ell\ell}$ is the di-lepton invariant mass, $\sqrt{s}$ is the center-of-mass energy of the $pp$ collision, 
$f_i$ are the parton distribution functions (PDFs), $x_{1,2} = \frac{M_{\ell\ell}}{\sqrt{s}} e^{\pm y}$ are the Bjorken 
variables and $y$ is the di-lepton rapidity (see \cite{Ellis:1991qj,Barger:1987nn,Bohm:1989pb}). The $C_u$ and $C_d$ 
coefficients are   given by (neglecting the $Z$ boson width $\Gamma_Z$)
\begin{widetext}
\onecolumngrid
\begin{eqnarray}
\label{eq2_nosimp}
&& C_u(\alpha_Y,\alpha_2,M_{\ell\ell}) = 576 \,s^{4}_W \,\alpha_2^2 \times \left[ \frac{4}{9} -  
\frac{\left( \frac{1}{2} - \frac{4}{3} s^2_W \right)  \left(- \frac{1}{2} + 2\, s^2_W \right)  }{3\, s^2_W\, (1-s^2_W) \left(1-\frac{m_Z^2}{M^2_{\ell\ell}}\right)} + 
\frac{ \left[\frac{1}{4} + \left( \frac{1}{2} - \frac{4}{3} s^2_W \right)^2 \right] \left[\frac{1}{4} + 
\left(\frac{1}{2} - 2\, s^2_W \right)^2 \right] }{16\, s^4_W\, (1-s^2_W)^2 \left(1-\frac{m_Z^2}{M^2_{\ell\ell}}\right)^2} 
\right]
 \,,\nonumber\\
&& C_d(\alpha_Y,\alpha_2,M_{\ell\ell}) = 576 \,s^{4}_W \,\alpha_2^2 \times \left[ \frac{1}{9} +  
\frac{\left( \frac{1}{2} - \frac{2}{3} s^2_W \right)  \left(\frac{1}{2} - 2\, s^2_W \right)  }{6\, s^2_W\, (1-s^2_W) \left(1-\frac{m_Z^2}{M^2_{\ell\ell}}\right)} + 
\frac{ \left[\frac{1}{4} + \left( \frac{1}{2} - \frac{2}{3} s^2_W \right)^2 \right] \left[\frac{1}{4} + 
\left(\frac{1}{2} - 2\, s^2_W \right)^2 \right] }{16\, s^4_W\, (1-s^2_W)^2 \left(1-\frac{m_Z^2}{M^2_{\ell\ell}}\right)^2} 
\right],
\end{eqnarray}
\end{widetext}
with $s^2_W = \alpha_Y / (\alpha_2 + \alpha_Y)$.
In the limit $M_{\ell\ell} \gg m_Z$,~\eqref{eq2_nosimp} simplifies to 
\begin{eqnarray}
\label{eq2}
&& C_u (\alpha_i) = 85 \alpha_Y^2 +6 \alpha_Y \alpha_2+ 9 \alpha_2^2 \,,\nonumber\\
&& C_d (\alpha_i) = 25 \alpha_Y^2 -6 \alpha_Y \alpha_2+ 9 \alpha_2^2 \,.
\end{eqnarray}
We note that this particularly simple form can be viewed as a result of the exchange of the hypercharge and $W_3$ bosons, 
which are effectively massless in this limit.

In the {\it leading log} approximation, the gauge couplings are to be understood as the running couplings taken at 
the scale $M_{\ell\ell}$, which controls the partonic center-of-mass energy of the Drell-Yan process, 
$\alpha_i \rightarrow \alpha_i (M_{\ell\ell})$. In this case,~\eqref{eq1} and~\eqref{eq2_nosimp} allow us to derive the variation
of $d\sigma/dM_{\ell\ell}$ under the change of the SM gauge couplings at the scale $\mu = M_{\ell\ell}$. 
For instance, taking $\sqrt{s}=8$ TeV and $\mu=1$ TeV, we find
\begin{equation}
\label{eq3}
\frac{ \delta \frac{d \sigma}{dM_{\ell\ell}} }{ \frac{d \sigma^{\mathrm{SM}}  }{dM_{\ell\ell}} } \Bigg\vert_{\mu = \rm 1\; TeV} \simeq
 0.84\, \frac{\delta \alpha_Y (\mu)}{\alpha^{\mathrm{SM}}_Y(\mu)} + 1.16 \, \frac{\delta \alpha_2 (\mu)}{\alpha^{\mathrm{SM}}_2(\mu)} \;,
\end{equation}
using NNPDF2.3~\cite{Ball:2012cx} at Next-to-Next-to-Leading-Order (NNLO). 
This illustrates that an $\mathcal{O}(1-10)\%$ precision measurement of $d\sigma/dM_{\ell\ell}$ can be translated 
into a bound on the variation of  $\delta\alpha_i/\alpha_i^{\mathrm{SM}} (\mu) $ of the same order.
In order to assess the sensitivity to new states, we define the ratio 
$ 
R(\mu,M_f, \Upsilon) =  \frac{d \sigma}{dM_{\ell\ell}}(\mu,M_f, \Upsilon )/\frac{d \sigma^{\mathrm{SM}}}{dM_{\ell\ell}}(\mu)\;
$
with $\Upsilon = N\,Y_f^2$ ($\Upsilon = N_d$) for the $SU(2)_L$ singlet (doublet) vector-like fermions.

We now turn to deriving constraints on the new EW states using the Drell-Yan 8 TeV (19.7 fb$^{-1}$) CMS analysis~\cite{CMS:2014jea}. 
The data include  $e^+e^- $ and $\mu^+ \mu^-$ lepton pairs with $M_{\ell\ell}$ up to 2 TeV, and the agreement between the SM prediction and 
experimental data is excellent, with the error bars well below 10\% all the way up to 1 TeV.
We use the CMS  data in the binned interval $M_{\ell\ell} \in$ [220 GeV,\, 2 TeV]. 
Since the bins have a sizable $\Delta M^i_{\ell\ell}$,
the ratio $R(\mu_i,M_f, \Upsilon)$ should be defined for each bin $i$ as an integral of $d \sigma/dM_{\ell\ell}$ over $\mu$ 
in the range $\Delta M^i_{\ell\ell}$, divided by the corresponding integral of $d \sigma^{\mathrm{SM}}/dM_{\ell\ell}$ over the same $\mu$ 
range. We  note that although~\eqref{eq1} is defined at LO in the hard scattering process, we include corrections up to NNLO 
through a $\mu$-dependent rescaling of the LO cross section.
To this end, we use the NNLO theoretical prediction for the Drell-Yan cross section~\cite{CMS:2014jea}
via {\sc FEWZ} 3.1~\cite{Gavin:2010az,Boughezal:2013cwa}.
We then perform a $\chi^2$ fit to the binned CMS 8 TeV  data using NNPDF2.3~\cite{Ball:2012cx} NNLO PDFs (and retaining finite $m_Z$ effects) with
\begin{equation}
\label{eq7}
\chi^2(M_f,\Upsilon) =  \sum_{\mu_i} \left(\frac{R^{\mathrm{exp}}_i - R(\mu_i,M_f,\Upsilon)}{\Delta R^{\mathrm{exp}}_i} \right)^2 \;.
\end{equation}
The data to theory ratio $R^{\mathrm{exp}}_i$ for each bin  as well as the corresponding uncertainty
$\Delta R^{\mathrm{exp}}_i$ are extracted from the CMS 8 TeV analysis~\cite{CMS:2014jea}. 
We show the resulting 1$\sigma$ and 2$\sigma$ bounds on $N\, Y_f^2$ vs $M_f$ and $N_d$ vs $M_f$ 
in Figure~\ref{fig1}-Left and~-Right, respectively.
The constraints on $N\,Y_f^2$ and $N_d$ are seen to be much stronger than those from EWPO. In Figure~\ref{fig1}-Left, we also show 
the bound from the SM gauge coupling perturbativity up to $\mu = 3$ TeV.

\vspace{2mm}

We go on to compute the projected LHC limits for $\sqrt{s}=14$ TeV with $300\,\,\mathrm{fb}^{-1}$, under the assumption that future experimental data
agree with the SM prediction ($R^{\mathrm{exp}}_i = 1$). The largely dominant source of systematic uncertainties in the Drell-Yan process 
for $M_{\ell\ell} \gg m_Z$ at $\sqrt{s}=14$ TeV is given by the PDF uncertainty~\cite{Alves:2014cda} (see also~\cite{Goertz:2016iwa}). 
We use the accurate estimate of the size of the PDF uncertainties in the Drell-Yan process from~\cite{Alves:2014cda}, adding this 
uncertainty in quadrature with the statistical uncertainty (which becomes dominant for $\mu \gtrsim 1$ TeV) to obtain a 
projected estimate of $\Delta R^{\mathrm{exp}}_{i}$. 
We then perform a $\chi^2$ fit using~\eqref{eq7}, and choosing 
$\mu_i = 1,\,1.5,\,2,\,2.5,\,3$ TeV. Our motivation for this choice is twofold: A substantial $M_{\ell\ell}$ separation between 
the $\mu_i$ bins implies weak correlations among them, allowing us to treat 
the bins in~\eqref{eq7} as uncorrelated\footnote{Our results agree quantitatively with those in~\cite{Alves:2014cda,Goertz:2016iwa}, 
where these weak correlations have been included.}. In addition, for values of $\mu$ beyond 3 TeV, the statistical uncertainty in the Drell-Yan 
process with $300\,\,\mathrm{fb}^{-1}$ becomes $\mathcal{O}(1)$. Our 1$\sigma$ and 2$\sigma$ projected 
bounds on $N\, Y_f^2$ vs $M_f$ ($N_d$ vs $M_f$) with $\sqrt{s}=14$ TeV and $300\,\,\mathrm{fb}^{-1}$ are 
shown in Figure~\ref{fig1}-Left (Figure~\ref{fig1}-Right).

Before moving on, a few comments are in order: \textsl{(i)} The models considered here are idealized in that all the new fields are taken to be identical. 
In reality one may expect a more complicated mass spectrum, and/or several states with different EW quantum numbers.
In this respect, the presence of states with other quantum numbers could affect dramatically the collider phenomenology of 
these models leading for example to   complicated final states which would obscure LHC searches. On the other hand, these states 
do not necessarily affect the SM gauge coupling running in any tangible way. 
This highlights the fact that our bounds rely on 
the cumulative effect of  new EW states and are fairly insensitive to further details of the model.
\noindent \textsl{(ii)} Employing the {\it leading log} approximation for the   SM gauge couplings is a source of 
theory uncertainty. However,   in our 8 TeV analysis this error is far subleading to the PDF, background and statistical 
uncertainties, and can  safely be neglected. In the 14 TeV analysis, this source of error is negligible for the $M_f$ mass range considered here, 
except for $M_f \to 1$ TeV where it is again largely subdominant to the PDF and statistical uncertainties.
\noindent \textsl{(iii)} Besides the neutral current Drell-Yan process $p p \to \gamma^*/Z^* \to \ell^+\ell^-$, the 
charged current process $p p \rightarrow W^* \rightarrow \ell \nu$ can also be used to constrain $\alpha_2$ via the 
transverse mass distribution $m_T$~\cite{Alves:2014cda}. This analysis is beyond the scope of our current study.
In the next Section, we show that the forward-backward asymmetry
can yield  useful information complementary   to that from   the invariant mass distribution.

\begin{widetext}
\onecolumngrid
\begin{figure}[h!]
\begin{center}
\includegraphics[width=0.49\textwidth]{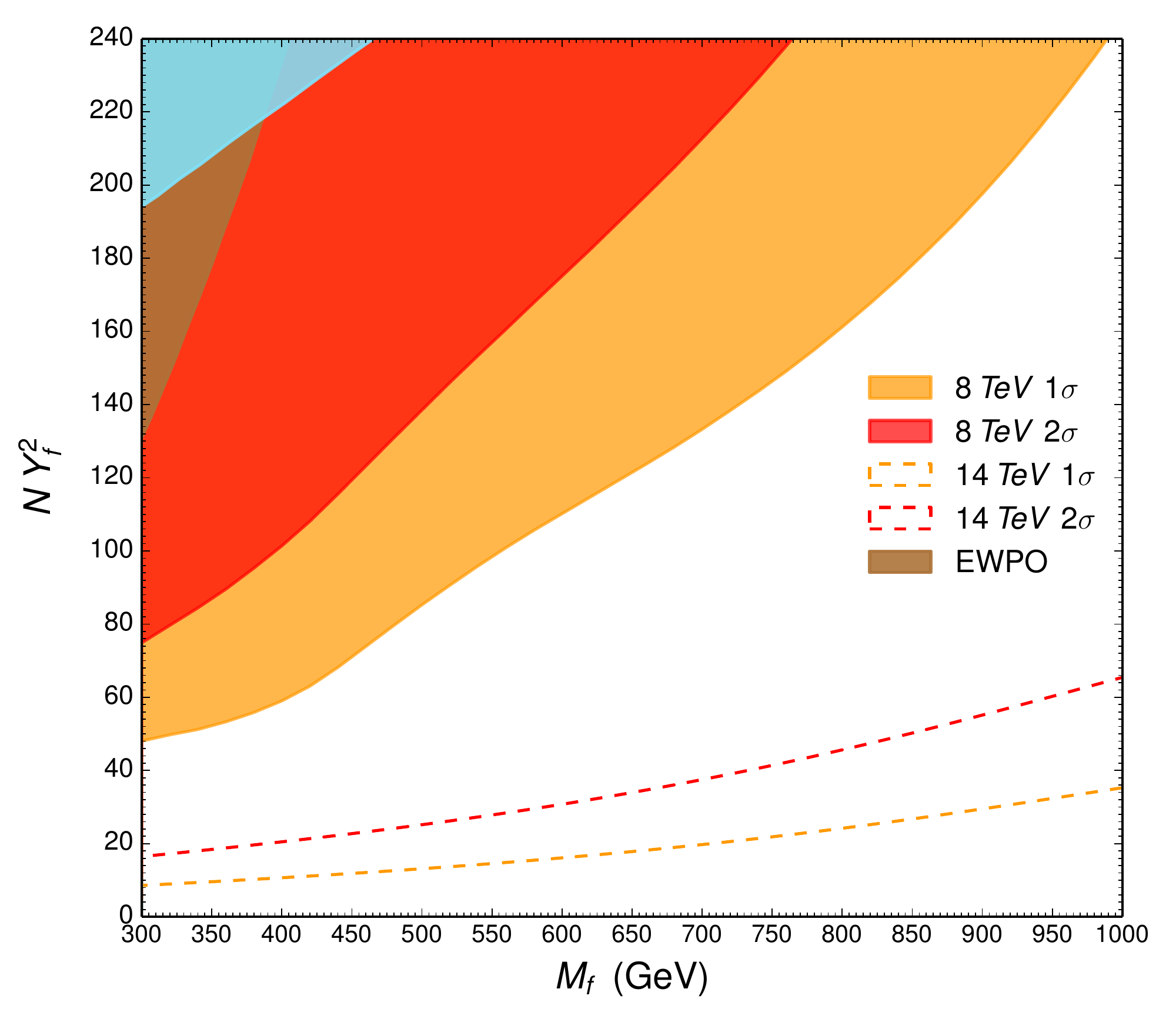}
\hspace{1mm}
\includegraphics[width=0.49\textwidth]{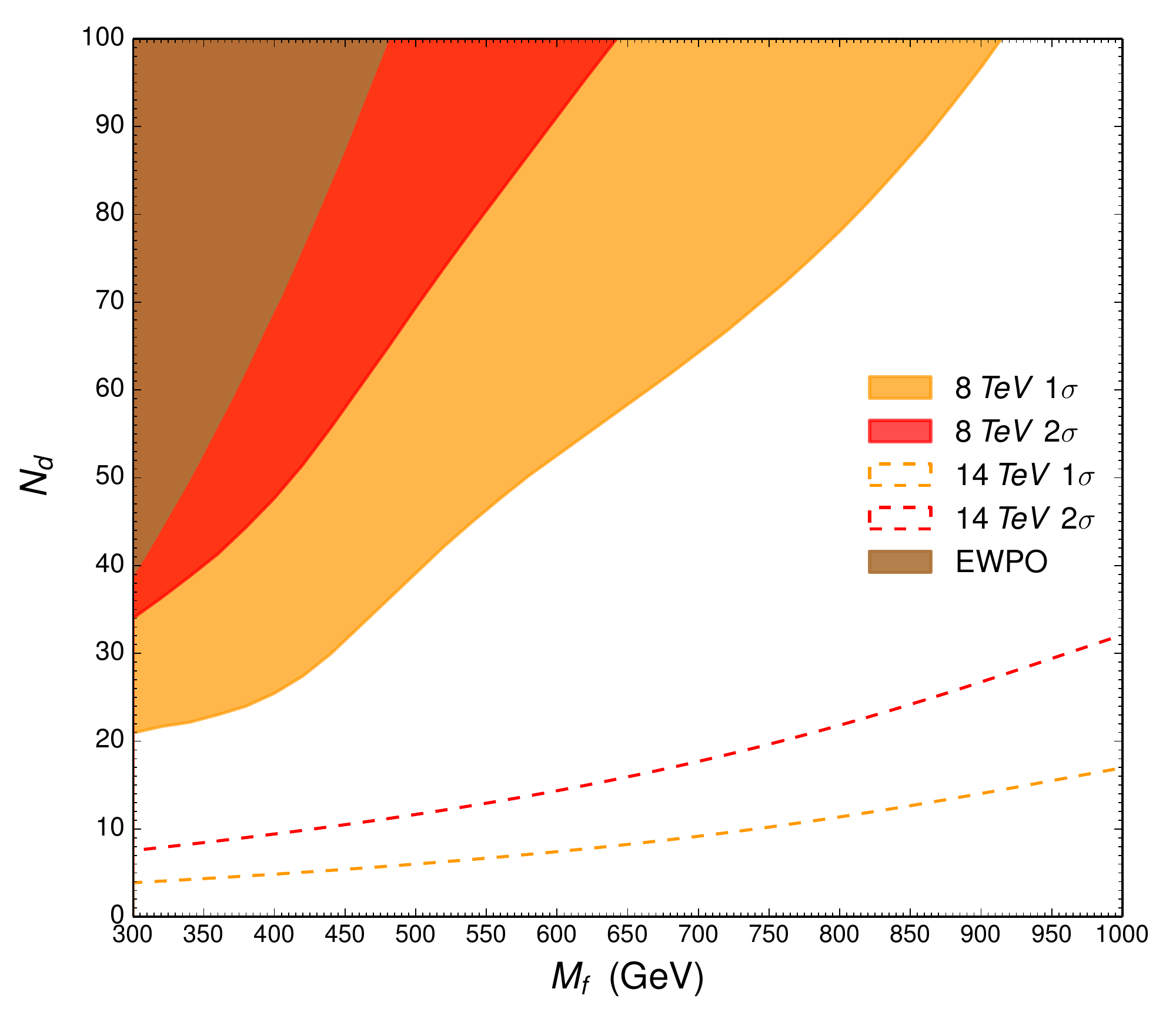}
\caption{\small {\it Left}: 8 TeV (19.7 fb$^{-1}$) Drell-Yan $1\sigma$ (orange) and $2\sigma$ (red) bounds on $SU(2)_L$ singlet vector-like fermions with hypercharge 
$Y_f$ and multiplicity $N$ vs mass $M_f$. The dashed red (orange) lines show the projected $2\sigma$ ($1\sigma$) bounds
from  14 TeV LHC (300 fb$^{-1}$). Also displayed are the EWPO constraint from the $\mathrm{Y}-$parameter $\mathrm{Y} < 2\times 10^{-3}$ 
(brown) and the $g_Y$ perturbativity constraint up to the scale $\mu \sim $ 3 TeV, $g_Y^2 (3 {\rm ~TeV}) < 4 \pi$ (light-blue).
{\it Right:} Same for $SU(2)_L$ doublets with zero hypercharge and multiplicity $N_d$. The EWPO constraint 
is due to the $\mathrm{W}$-parameter $\mathrm{W} < 1\times 10^{-3}$.
\label{fig1}}
\end{center}
\end{figure}
\end{widetext}

\noindent \textbf{Di-Lepton Forward-Backward Asymmetry}.~In addition to the di-lepton invariant mass distribution $d\sigma/dM_{\ell\ell}$, an observable which can be useful in constraining 
the    gauge couplings is the forward-backward asymmetry $A_{\rm FB}$. At the parton level, in the di-lepton center-of-mass 
frame, one has~\cite{Aad:2009wy}
\begin{eqnarray}
\frac{d \sigma (q \bar q \rightarrow \ell^+ \ell^-)}{d \cos \theta^*} &=&
 \frac{1}{3} \left[ (1+ \cos^2 \theta^*)\;  F_0^q(M_{\ell \ell}^2) \right. \nonumber\\
  &+& \left. 2 \cos \theta^* \; F_1^q(M_{\ell \ell}^2)    \right] \;,
\end{eqnarray}
where $\theta^* $ is the angle between the quark and $\ell^-$ momenta.
For $ M_{\ell \ell} \gg m_Z$, the form--factors $F_{0,1}^q(M_{\ell \ell}^2)$ read
\begin{eqnarray}
&& F_0^u = \frac{\pi \, C_u(\alpha_i)}{1152 \,M_{\ell\ell}^2} ~,~
 F_1^u = \frac{\pi\,(15 \alpha_Y^2 + 2 \alpha_Y \alpha_2 + 3 \alpha_2^2 )}{384\, M_{\ell\ell}^2}\, ,\nonumber\\
&& F_0^d = \frac{\pi \, C_d(\alpha_i)}{1152 \,M_{\ell\ell}^2} ~,~
 F_1^d = \frac{\pi\,(3 \alpha_Y^2 - 2 \alpha_Y \alpha_2 + 3 \alpha_2^2 )}{384 M_{\ell\ell}^2} \,. \nonumber
\end{eqnarray}
The forward-backward asymmetry $A_{\rm FB}$ is defined by 
\begin{eqnarray}
A_{\rm FB}^q (M_{\ell \ell}^2) = \frac{\sigma (\cos\theta^* >0) - \sigma (\cos\theta^* <0)}{ 
\sigma (\cos\theta^* >0) + \sigma (\cos\theta^* <0)}= \frac{3 F_1^q(M_{\ell \ell}^2)}{ 4 F_0^q(M_{\ell \ell}^2)} \;. \nonumber
\end{eqnarray}
It is  proportional to the quark and lepton axial couplings. 
In order to understand the sensitivity of  $A_{\rm FB}$  to the gauge couplings, let us consider the parton level variations
\begin{eqnarray}
&&\frac{\delta A_{\rm FB}^u}{A_{\rm FB}^u  } \Bigg\vert_{\mu=\rm 1\; TeV} \simeq
0.27 \left(  - \frac{\delta \alpha_Y (\mu)}{ \alpha_Y^{\rm SM}(\mu)}  + \frac{\delta \alpha_2 (\mu)}{ \alpha_2^{\rm SM}(\mu)} \right)
\;,~ ~\nonumber\\ 
&&\frac{\delta A_{\rm FB}^d}{ A_{\rm FB}^d  } \Bigg\vert_{\mu=\rm 1\; TeV} \simeq
0.35 \left(  -\frac{\delta \alpha_Y (\mu)}{ \alpha_Y^{\rm SM}(\mu)}  + \frac{\delta \alpha_2(\mu) }{ \alpha_2^{\rm SM}(\mu)} \right)\, ,
\end{eqnarray}
where in the SM $A_{\rm FB}^{u,d} \simeq 0.6$ at $\mu = 1$ TeV.  
Comparing this to~\eqref{eq3}, we immediately see that $A_{\rm FB}$ can provide information complementary to  that from  $d\sigma/dM_{\ell\ell}$.
Although the asymmetry is less sensitive to the variation of the couplings than the  
differential cross section, it can still be a useful observable, having the 
advantage that many systematic and QCD uncertainties cancel in $A_{\rm FB}$. 
It has however its own problematic issues, above all the reconstruction of the quark 
direction, which is reliable for large rapidities only \cite{Aad:2009wy}. Current measurements at 8 TeV LHC~\cite{Khachatryan:2016yte,Aad:2015uau}
contain substantial uncertainties of about 25\% at large $M_{\ell\ell}\sim 1$ TeV 
which make the constraints inferior to those from $d\sigma/dM_{\ell\ell}$.
Nevertheless, $A_{\rm FB}$ can be a useful complementary observable in 
the future LHC runs.

\vspace{2mm}

\noindent \textbf{Comments on Direct Searches}. In what follows, we provide a more detailed discussion of
collider searches for new EW   states. We first focus on vector-like (VL) leptons:  
for $SU(2)_L$ singlet VL leptons with SM-like charges, LEP searches  
yield a lower bound $M_f \gtrsim 100$ GeV~\cite{Achard:2001qw}. 
For a non-zero mixing between these and the SM leptons, the bounds from LEP2 and LHC searches can be 
extended up to $M_f \gtrsim 200$ GeV~\cite{Falkowski:2013jya,Aad:2015dha} and higher in some cases. In this case, there are also relevant constraints 
from  the anomalous magnetic moment of the muon~\cite{Mery:1989dx}.
VL leptons with exotic, non-SM charges can only decay via higher-dimensional operators. If these are stable on detector scales, 
the lower limit on their masses is $M_f \gtrsim 700 \ldots 800$~GeV for $|Q| = 2 e$ to $|Q| = 8 e$~\cite{Chatrchyan:2013oca}.
However, if they decay promptly to multi-lepton final states and/or jets, no general bounds exist as these are highly 
model dependent and would require a dedicated analysis.

For VL quarks, the constraints on their masses are in general significantly stronger, although still model dependent. 
For VL partners of $t$ or $b$ quarks, as well as for VL quarks with exotic charges 5/3 and 8/3, 
the LHC Run 1 limit is $M_f \gtrsim 800 \ldots 1000$ GeV~\cite{Aad:2015mba,Matsedonskyi:2014lla}. This assumes that they decay into $t + X$ and/or $b + X$ (with $X$ unspecified), while the 
 precise limits depend on the corresponding  branching ratios.


\vspace{-3mm}

\section{Conclusions}

\vspace{-2mm}

Current and future Drell-Yan measurements at the LHC place important constraints on possible new states with EW quantum numbers.
Such states modify the 
 RG running of the SM gauge couplings at high energies 
thereby affecting the Drell-Yan production rates. 
The resulting  bounds are fairly insensitive to the details of 
the mass spectrum or interactions among these states and therefore are complementary to the direct search constraints. 
We have shown that both the di-lepton differential $d\sigma/dM_{\ell\ell}$ distribution and 
the forward-backward asymmetry $A_{\rm FB}$ can place meaningful constraints on the existence of new EW states, and that 
the LHC sensitivity of $p p \to \ell^+\ell^-$   to the  new states far surpasses that of the current EW precision measurements.

\begin{center}
\textbf{Acknowledgements} 
\end{center}
\vspace{-2mm}
\begin{acknowledgments}
We thank Adam Falkowski for helpful communications. C.G. and O.L. acknowledge support from the Academy of Finland, project ``The Higgs and the Cosmos''.
J.M.N. is supported by the People Programme (Marie Curie Actions) of the European Union Seventh 
Framework Programme (FP7/2007-2013) under REA grant agreement PIEF-GA-2013-625809.
\end{acknowledgments}


\begin{thebibliography}{99}
  
\bibitem{Barbieri:2004qk}
  R.~Barbieri, A.~Pomarol, R.~Rattazzi and A.~Strumia,
  Nucl.\ Phys.\ B {\bf 703}, 127 (2004)
  [hep-ph/0405040].
 
 
\bibitem{Achard:2001qw}
  P.~Achard {\it et al.} [L3 Collaboration],
  Phys.\ Lett.\ B {\bf 517} (2001) 75
  [hep-ex/0107015].
  
\bibitem{Falkowski:2013jya}
  A.~Falkowski, D.~M.~Straub and A.~Vicente,
  JHEP {\bf 1405} (2014) 092
  [arXiv:1312.5329 [hep-ph]].
  
\bibitem{Aad:2015dha}
  G.~Aad {\it et al.} [ATLAS Collaboration],
  JHEP {\bf 1509} (2015) 108
  [arXiv:1506.01291 [hep-ex]].
 
\bibitem{Djouadi:2016eyy}
  A.~Djouadi, J.~Ellis, R.~Godbole and J.~Quevillon,
  JHEP {\bf 1603} (2016) 205
  [arXiv:1601.03696 [hep-ph]].

\bibitem{Alves:2014cda}
  D.~S.~M.~Alves, J.~Galloway, J.~T.~Ruderman and J.~R.~Walsh,
  JHEP {\bf 1502} (2015) 007
  [arXiv:1410.6810 [hep-ph]].

  
\bibitem{Drell:1970wh} 
  S.~D.~Drell and T.~M.~Yan,
  Phys.\ Rev.\ Lett.\  {\bf 25}, 316 (1970).
  
  
  \bibitem{Ellis:1991qj}
  R.~K.~Ellis, W.~J.~Stirling and B.~R.~Webber,
  Camb.\ Monogr.\ Part.\ Phys.\ Nucl.\ Phys.\ Cosmol.\  {\bf 8} (1996) 1.
  
  \bibitem{Barger:1987nn}
  V.~D.~Barger and R.~J.~N.~Phillips,
  REDWOOD CITY, USA: ADDISON-WESLEY (1987) 592 P. (FRONTIERS IN PHYSICS, 71).
  
  \bibitem{Bohm:1989pb}
  M.~Bohm {\it et al.},
  CERN-TH-5536-89.
  
 

\bibitem{Ball:2012cx} 
  R.~D.~Ball {\it et al.},
  Nucl.\ Phys.\ B {\bf 867}, 244 (2013)
  [arXiv:1207.1303 [hep-ph]].
  
    \bibitem{CMS:2014jea}
  V.~Khachatryan {\it et al.} [CMS Collaboration],
  Eur.\ Phys.\ J.\ C {\bf 75} (2015) 4,  147
  [arXiv:1412.1115 [hep-ex]].

  
\bibitem{Gavin:2010az} 
  R.~Gavin, Y.~Li, F.~Petriello and S.~Quackenbush,
  Comput.\ Phys.\ Commun.\  {\bf 182}, 2388 (2011)
  [arXiv:1011.3540 [hep-ph]].  
  
\bibitem{Boughezal:2013cwa} 
  R.~Boughezal, Y.~Li and F.~Petriello,
  Phys.\ Rev.\ D {\bf 89}, no. 3, 034030 (2014)
  [arXiv:1312.3972 [hep-ph]].
  
\bibitem{Goertz:2016iwa} 
  F.~Goertz, A.~Katz, M.~Son and A.~Urbano,
  JHEP {\bf 1607}, 136 (2016)
  [arXiv:1602.04801 [hep-ph]]. 
 
  
\bibitem{Aad:2009wy} 
  G.~Aad {\it et al.} [ATLAS Collaboration],
  arXiv:0901.0512 [hep-ex], p.815.
  
 
\bibitem{Khachatryan:2016yte}
  V.~Khachatryan {\it et al.} [CMS Collaboration],
  Eur.\ Phys.\ J.\ C {\bf 76} (2016) no.6,  325
  [arXiv:1601.04768 [hep-ex]].

\bibitem{Aad:2015uau} 
  G.~Aad {\it et al.} [ATLAS Collaboration],
  JHEP {\bf 1509}, 049 (2015) 
  [arXiv:1503.03709 [hep-ex]].
  
  
 \bibitem{Mery:1989dx} 
  P.~Mery, S.~E.~Moubarik, M.~Perrottet and F.~M.~Renard,
  Z.\ Phys.\ C {\bf 46}, 229 (1990).
  
  
  
\bibitem{Chatrchyan:2013oca}
  S.~Chatrchyan {\it et al.} [CMS Collaboration],
  JHEP {\bf 1307} (2013) 122
  [arXiv:1305.0491 [hep-ex]].

 
\bibitem{Aad:2015mba}
  G.~Aad {\it et al.} [ATLAS Collaboration],
  Phys.\ Rev.\ D {\bf 91} (2015) no.11,  112011
  [arXiv:1503.05425 [hep-ex]];
  %
  G.~Aad {\it et al.} [ATLAS Collaboration],
  JHEP {\bf 1508} (2015) 105
  [arXiv:1505.04306 [hep-ex]];
 %
  G.~Aad {\it et al.} [ATLAS Collaboration],
  Phys.\ Rev.\ D {\bf 92} (2015) no.11,  112007
  [arXiv:1509.04261 [hep-ex]].
  
\bibitem{Matsedonskyi:2014lla}
  O.~Matsedonskyi, F.~Riva and T.~Vantalon,
  JHEP {\bf 1404} (2014) 059
  [arXiv:1401.3740 [hep-ph]].

%
%
%
%
   
  
 
  
\end{thebibliography}
\end{document}